\documentclass[runningheads]{llncs}
\usepackage{amssymb}

\newtheorem{insights}{RQ}

\usepackage{longfbox}
\usepackage{graphicx}
\usepackage[T1]{fontenc}
\usepackage[edges]{forest}
\usepackage{amsmath}
\usepackage{xspace}
\usepackage{booktabs}
\usepackage{makecell}

\usepackage{todonotes}
\setuptodonotes{inline, color=blue!30}
\makeatletter
\let\wfs@comment@comment\comment
\let\comment\@undefined
\usepackage[commandnameprefix=ifneeded,commentmarkup=todo, draft]{changes}

\usepackage{hyperref}
\hypersetup{
    colorlinks=true,
    linkcolor=blue,
    filecolor=magenta,      
    urlcolor=cyan,
    pdfpagemode=FullScreen,
}

\usepackage{changepage}   
\usepackage{calc}         

\usepackage{acro}
\acsetup{single}

\usepackage[font=small]{caption}
\usepackage{tabularx}
\usepackage{appendix}
\usepackage{enumitem}

\begin{document}

\title{Bittensor Protocol: The Bitcoin in Decentralized Artificial Intelligence? A Critical and Empirical Analysis}

\author{Elizabeth Lui \and Jiahao Sun}
\institute{
FLock.io
}

\titlerunning{Bittensor: Bitcoin of Decentralized AI?}
\authorrunning{Lui and Sun}

\maketitle
\begin{abstract}

This paper investigates whether Bittensor can be considered the “Bitcoin of decentralized Artificial Intelligence” by directly comparing its tokenomics, decentralization properties, consensus mechanism, and incentive structure against those of Bitcoin. Leveraging on‐chain data from all 64 active Bittensor subnets, we first document considerable concentration in both stake and rewards. We further show that rewards are overwhelmingly driven by stake, highlighting a clear misalignment between quality and compensation. As a remedy, we put forward a series of two-pronged protocol-level interventions. For incentive realignment, our proposed solutions include performance‐weighted emission split, composite scoring, and a trust‐bonus multiplier. As for mitigating security vulnerability due to stake concentration, we propose and empirically validate stake cap at the 88nd percentile, which elevates the median coalition size required for a 51\% attack and remains robust across daily, weekly, and monthly snapshots.

\end{abstract}

\section{Introduction}

Bitcoin’s emergence in 2009 introduced a decentralized digital currency secured by cryptography and game‑theoretic incentives\cite{nakamoto2008bitcoin}. Through its proof‑of‑work (PoW) consensus and strictly limited supply, Bitcoin created a trustless system where value is secured by computational cost and scarcity. Recently, protocols like Bittensor aim to apply similar principles to a new domain—decentralized Artificial Intelligence (deAI)~\cite{wang2024sok}. Bittensor proposes a blockchain‑based network where participants are rewarded for contributing machine learning models and evaluating others’ models, in effect creating a marketplace for “collectively produced” Artificial Intelligence (AI). In fact, Bittensor explicitly references Bitcoin in its whitepaper:  
\begin{quote}
  “Bittensor is like Bitcoin in many ways. It has a transferrable and censorship resistant token, TAO, which runs on a 24/7 decentralized blockchain substrate which is auditable and transparent. Bittensor is also run by miners, like Bitcoin, who can exist globally and anonymously.”\cite{bittensor_about}
\end{quote}

Moreover, Bittensor is today the largest deAI project by market capitalization—currently around \$2.12 billion\cite{coinmarketcap_bittensor}—making it a natural case study for understanding how Bitcoin‑inspired design can be extended into the AI realm.  This is particularly important as  deAI frameworks are critical for preserving data privacy, improving system robustness, and preventing concentration of model ownership in a handful of large technology platforms \cite{wang2024sok}.


The contributions of this paper are as follows:

\begin{itemize}[leftmargin=*, labelsep=0.5em]
  \item \textbf{Longitudinal empirical study.}  We assemble and analyze nearly two years of staking, reward and performance data across all active Bittensor subnets — the first of such large‑scale measurement, to the best of our knowledge.
  \item \textbf{Decentralization metrics.} We compute Gini coefficients, Herfindahl--Hirschman \\Index (HHI), and top‑1\% share for both stake and rewards. We relate these to subnet size and age, and show how concentration varies dramatically by subnet.
  \item \textbf{51\%‑attack vulnerability.}  We calculate, per subnet, the fraction of wallets needed to command a majority of stake, revealing that many subnets can be trivially compromised by coalitions of just 1–2\% of participants.
  \item \textbf{Performance–economics correlations.}  By aggregating performance metrics for miners and validators respectively and correlating them with stake and reward, we show that while stake strongly predicts earnings, performance scores are only weakly rewarded — pointing to opportunities for incentive redesign.
  \item \textbf{Protocol interventions.} We propose two complementary classes of protocol‐level refinements:
  \begin{itemize}
    \item \emph{Incentive realignment:} performance‐weighted emission split, composite scoring, and a trust‐bonus multiplier to strengthen the perf$~\rightarrow$reward link;
    \item \emph{Security mitigation:} stake cap at the 88nd percentile, concave stake transforms, and reward demurrage to raise the 51\% attack threshold.
  \end{itemize}
  We empirically validate each scheme’s effect on perf$~\rightarrow$reward and stake
  $~\rightarrow$reward correlations, and demonstrate the security–penalty trade‑offs and temporal robustness of our mitigation strategies across daily, weekly, and monthly snapshots.
\end{itemize}

\section{Background}
\subsection{Bitcoin’s Tokenomics and Decentralization}

\begin{table}[h]
  \centering
  \small
  \begin{tabularx}{\columnwidth}{p{3cm}X X}
    \toprule
    \textbf{Attribute} & \textbf{Bitcoin} & \textbf{Bittensor} \\
    \midrule
    Native Token
      & BTC, capped at 21 million; halvings every $\sim$4 years  
      & TAO, capped at 21 million; halvings on a similar schedule \\[6pt]
    Consensus
      & PoW: longest valid chain by majority hashpower (attack requires $>50\%$)
      & \textbf{Yuma Consensus}: subjective‐utility ranking by validators, stake‐weighted median clipping, rewards to miners \& validators \\[6pt]
    Roles
      & Miners only
      & Miners, Validators, Delegators, plus Subnet Owners \\[6pt]
    Reward Split
      & All block subsidy to miners
      & 41\% to miners, 41\% to validators, 18\% to subnet owners \\[6pt]
    Security Assumption
      & Honest majority of hashing power
      & Honest majority stake \emph{and} utility, resilient to minority collusion \\[6pt]
    Application
      & Decentralized currency and store of value
      & Decentralized market for AI models and evaluations \\ 
    \bottomrule
  \end{tabularx}
  \caption{High‐level comparison of Bitcoin and Bittensor.}
  \label{tab:btc_vs_btao}
\end{table}

Bitcoin is founded on an economic model in which digital scarcity is enforced by code. The Bitcoin protocol limits the total supply to 21 million BTC, introduced at a diminishing rate through mining rewards that halve approximately every four years (the “halving” schedule) \cite{nakamoto2008bitcoin}. This predictable stock‑to‑flow schedule instills scarcity, a key pillar of Bitcoin’s value proposition \cite{cheah2015speculative,gandal2018price}. Every BTC is costly to obtain, and the fixed cap signals long‑term rarity, similar to gold’s finite supply \cite{yermack2013economic,bohme2015bitcoin}.

Bitcoin’s consensus mechanism further underpins its security and decentralization. PoW mining requires nodes to perform energy‑intensive computations to solve cryptographic puzzles. The network accepts the longest valid chain as truth—an attacker would need more than 50~\% of global hashing power to rewrite history. As long as a majority of miners act honestly, the longest chain represents the heaviest invested work and thus the valid ledger \cite{nakamoto2008bitcoin,bonneau2015sok,eyal2014majority}.

\subsection{Bittensor’s Vision and Architecture}

As denoted in Table~\ref{tab:btc_vs_btao}, Bittensor extends Bitcoin’s paradigm of decentralized value exchange to AI compute and evaluation.  Like Bitcoin, it uses a capped supply of 21 million tokens (TAO) with periodic halvings \cite{bittensor_docs}, but instead of PoW it relies on a multi‐role, stake‐based “subjective‐utility” consensus.  Each block mints TAO which is split 41~\% to AI “miners” (who run inference), 41~\% to “validators” (who score that work), and 18~\% to the subnet creator.  Subnets are independent communities—each defined by an off‐chain incentive mechanism specifying the mining task, validation protocol, and emission rules—and their relative share of network emissions is governed by the 64 largest validators (the “root network”).  

On‐chain, the Yuma Consensus (YC) algorithm aggregates validators’ score vectors into final rewards: it computes a stake‐weighted median benchmark, clips outlier weights, and allocates miner emissions proportional to the clipped aggregate, while validator rewards accrue via exponentially smoothed bonds that penalize deviations from consensus.  This design seeks to incentivize honest, high‐quality participation and resists collusion up to a 50~\% validator stake.  

\section{Bittensor's System Overview}

\begin{figure*}[t]
  \begin{adjustwidth}{-\dimexpr1in+\hoffset+\oddsidemargin\relax}{0pt}
    \includegraphics[
      width=\paperwidth,
      trim=0 35 0 30,
      clip
    ]{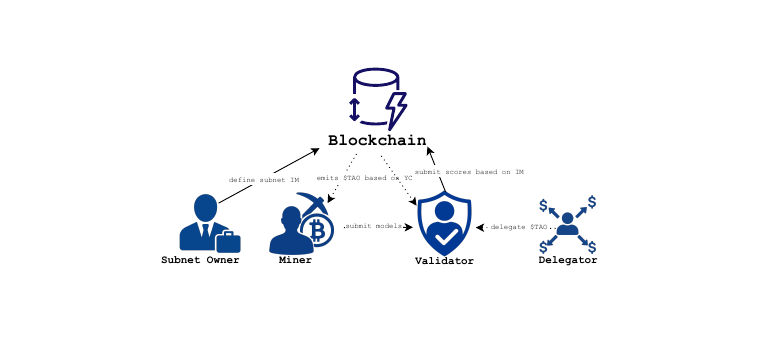}
  \end{adjustwidth}
  \caption{System Design Overview of Bittensor.}
  \label{fig:bittensor}
\end{figure*}

Before diving into our empirical study, we first give a high‐level overview of how Bittensor operates, from subnet creation through mining, validation, and delegation. Figure~\ref{fig:bittensor} shows an overview of Bittensor's system design.

\subsection{Blockchain Layer}
Subtensor is the blockchain protocol that underlies the Bittensor network, effectively serving as its “mainnet.” Built on a Substrate framework, Subtensor is responsible for:

\begin{itemize}
  \item Recording all staking, reward, and performance events;
  \item Hosting the YC logic on‑chain;
  \item Managing subnets, wallets, and token transfers of TAO.
\end{itemize}

\subsection{Subnets as AI Marketplaces}  
Bittensor is composed of a pool of \emph{subnets}, each of which is a self‐contained, incentive‐driven marketplace for a particular AI commodity (e.g.\ text prompts, translation, storage).  A subnet creator defines an off‐chain “incentive mechanism” (IM) that specifies:
\begin{itemize}
  \item \textbf{Miner Task:} the work miners must perform (e.g.\ respond to a prompt, store a file).
  \item \textbf{Validator Protocol:} how validators query miners and score their outputs, yielding a numerical weight vector per epoch.
\end{itemize}
Once deployed, a subnet is launched on the Subtensor blockchain and becomes discoverable to participants.

\subsection{Participant Roles and Workflow}

Bittensor participants interact in a coordinated cycle of task definition, execution, evaluation, and staking:

\begin{description}
  \item[Subnet Creator]  
    Defines the purpose and rules of a subnet (the “incentive mechanism”), deploys it on the blockchain, and remains credited with a fixed share of emissions.

  \item[Miner]  
    Competes to perform the subnet’s AI task (e.g.\ answering prompts or storing data).  Successful responses earn TAO proportional to their aggregated quality scores.

  \item[Validator]  
    Assesses miners’ work according to the subnet’s rules, assigning scores that are later combined via YC.  Validators’ own rewards depend on how closely their scores align with the stake‐weighted majority.

  \item[Delegator]  
    Stakes TAO on one or more validators to share in their earnings.  This lets TAO holders support validators whose evaluations they trust and helps decentralize network security.
\end{description}

\subsection{On‑Chain Coordination and Emissions}
At the end of each \emph{tempo} (360 blocks), Subtensor collects the latest weight vectors from all validators in each subnet and feeds them into the YC module.  YC applies stake‐weighted clipping to produce:
\begin{itemize}
  \item \textbf{Miner Emissions:} allocated proportionally to each miner’s consensus‐clipped aggregate score.
  \item \textbf{Validator Emissions:} allocated based on exponentially smoothed “bonds” that reward validators for alignment with the stake‐weighted median.
  \item \textbf{Subnet Owner Share:} a fixed 18\% of each subnet’s block emission goes directly to the creator.
   \item \textbf{Delegator Share:} extract their share of validator emissions in proportion to their staked TAO.
\end{itemize}

This architecture—multiple subnets, role‐based tasks, stakeholder delegation, and on‐chain resolution via YC—forms the backbone of Bittensor’s decentralized AI marketplace.  In the next sections, we quantify how stake, reward, and performance flows through this system and assess its resilience to concentration risks. See ~\ref{appendix:reward} for detailed discussion of the tokenomics of Bittensor. 

\section{Empirical Analysis of Bittensor’s Network}

\subsection{Research Questions \& Decentralization Framework}

Our central question is: \emph{Can Bittensor achieve Bitcoin‐grade decentralization in its AI‐market consensus?}  To answer this, we combine empirical network analysis with well–established decentralization metrics, asking:

\begin{enumerate}
  \item \textbf{Stake\,vs.\ Reward Dynamics.}  
    \emph{How tightly linked are staking power and reward share in Bittensor, for both miners and validators?}  In Bitcoin's PoW, block rewards scale (roughly) linearly with hash‐rate~\cite{eyal2014majority}—does Bittensor’s model exhibit similar or more skewed dynamics?
    
  \item \textbf{Decentralization Benchmarks.}  
    \emph{How does Bittensor’s stake concentration compare, subnet by subnet, to Bitcoin’s miner concentration?}  
    We compute Gini Coefficients, HHI and \emph{top-$k$\% share} exactly as one would for hashing‐power pools in Bitcoin\cite{gervais2016security}, and then ask whether Bittensor subnets approach Bitcoin’s historical decentralization levels (e.g.\ no single pool>50\%). See ~\ref{appendix:gini} for definitions of Gini Coefficients and HHI. 

  \item \textbf{51\%‐Attack Resilience.}  
    \emph{What fraction of wallets must collude to gain majority control in each subnet?}  
    Analogous to “what fraction of hash‐power” in Bitcoin~\cite{decker2013information}, we compute the minimal coalition size across subnets, and test whether smaller/newer subnets are inherently more vulnerable.
\end{enumerate}




\subsection{Measurement Setup}

All on‑chain data used in this work were obtained via the Taostats Metagraph History API\cite{taostats_metagraph-history}, which returns a chronological series of “events” representing daily snapshots of each wallet’s participation as a miner or validator on Bittensor.  Each event includes:
\begin{itemize}
  \item \textbf{Timestamp \& Block Number}: when the snapshot was taken.
  \item \textbf{Stake}: the amount of TAO staked by the wallet in that subnet.
  \item \textbf{Reward}: the daily emissions received by that wallet.
  \item \textbf{Performance Metrics}: \texttt{trust} for miners; \texttt{validator trust} for validators.
\end{itemize}

In total, we retrieved 6,664,830 events spanning from 2023-03-20 till 2025-02-12. Over this period, the data encompass a total of 121,567 unique wallets in all 64 subnets. The participants are comprised of 106,839 miners and 37,642 validators. Note that in a given subnet, a wallet can only either be a miner or a validator, but its roles can be different across different subnets.

It is noteworthy that on 2025-02-13, Bittensor introduced the DynamicTAO (dTAO) upgrade\cite{bittensor_dtao2025}, which replaces the centralized root‐network valuation mechanism with a fully distributed, stake‐based subnet valuation model.  To maintain a consistent protocol environment for our analysis, we have excluded all data after this upgrade.  We however anticipate that dTAO will materially affect the decentralization dynamics of Bittensor, motivating future empirical study of its impacts.

\section{Findings}

\begin{figure}[h]
  \centering
  \includegraphics[width=0.48\textwidth,trim=0 100 0 100,clip]{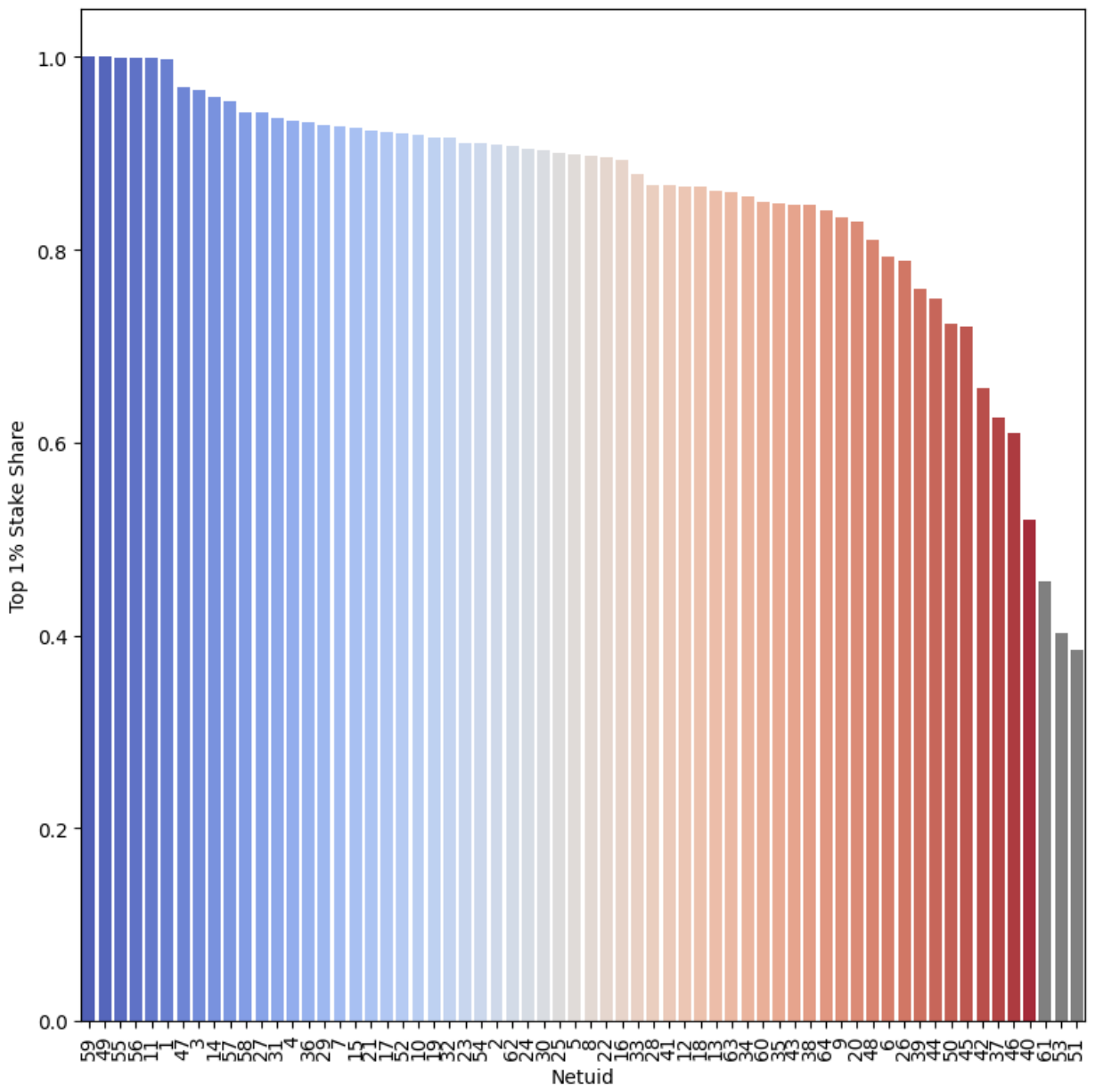}
  \hfill
  \includegraphics[width=0.48\textwidth,trim=0 100 0 100,clip]{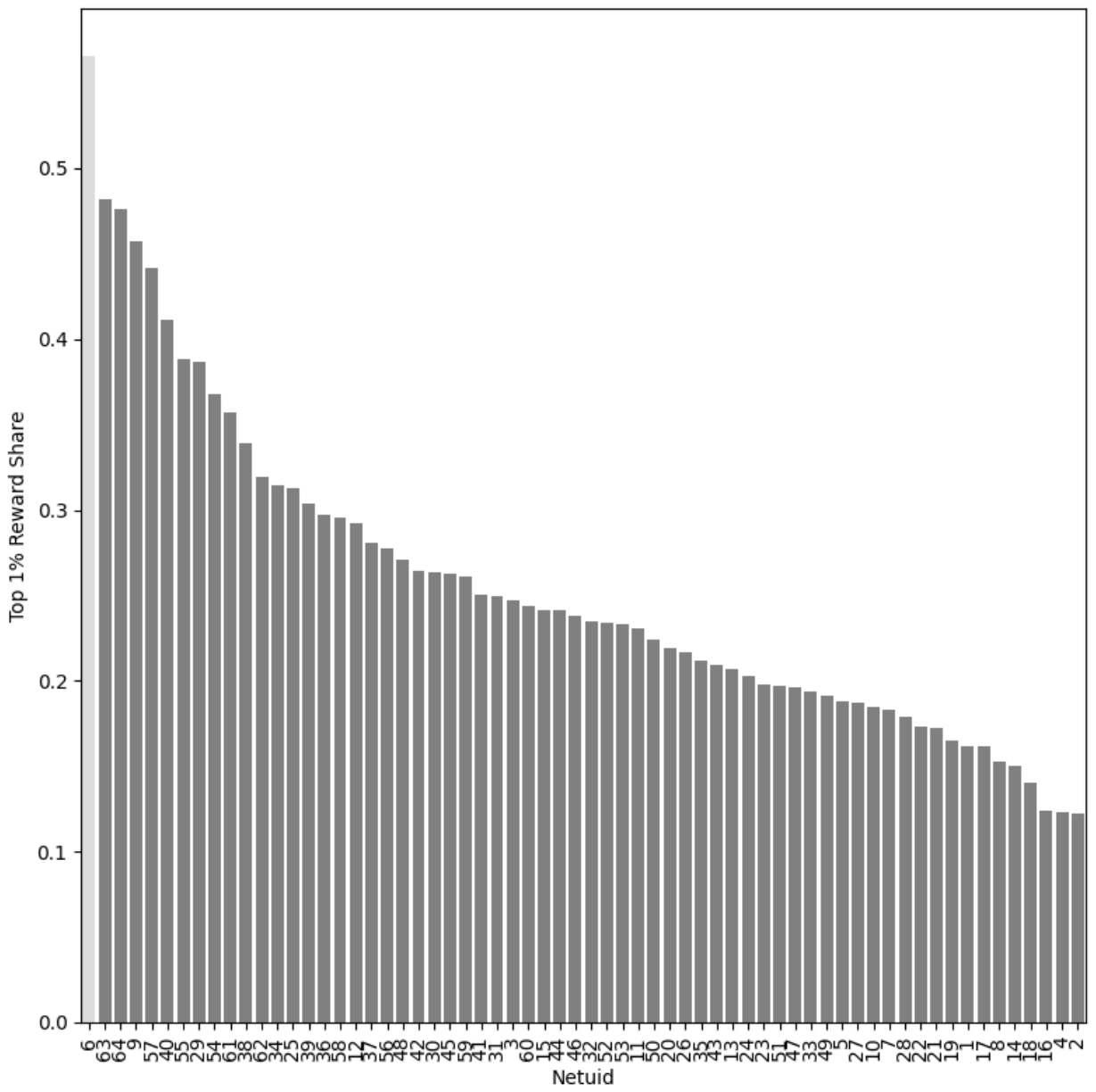}
  \caption{%
    \textbf{Left:} Top 1\% Share of Total Stake by Subnet.  
    \textbf{Right:} Top 1\% Share of Total Reward by Subnet.  
    Subnets are sorted left→right by decreasing concentration in each panel.%
  }
  \label{fig:top1_parallel}
\end{figure}

\begin{table}[h]
  \centering
  \small
  \begin{tabular}{lcccccc}
    \toprule
    & \multicolumn{3}{c}{\bfseries Stake Metrics} 
    & \multicolumn{3}{c}{\bfseries Reward Metrics} \\[3pt]
    \cmidrule(r){2-4}\cmidrule(l){5-7}
    \textbf{Metric} 
      & Mean & Median & Range 
      & Mean & Median & Range \\
    \midrule
    Overall Gini           & 0.9825 & 0.9879       & [0.7275,\,0.9984]
                           & 0.7561 & 0.9267       & [0.1958,\,0.9933] \\
    Overall HHI            & 0.1377 & 0.0886       & [0.0335,\,0.9997]
                           & 0.0623 & 0.0488       & [0.0003,\,0.6304] \\
    Top 1\% Share          & 0.8529 & 0.8984       & [0.3848,\,0.99997]
                           & 0.2557 & 0.2362       & [0.1221,\,0.5653] \\[4pt]
    \multicolumn{7}{l}{\itshape By Role} \\[2pt]
    Miner Gini             & 0.9880 & 0.9952       & [0.7275,\,0.9984]
                           & 0.5332 & 0.5030       & [0.1958,\,0.9480] \\
    Miner HHI              & 0.1839 & 0.1034       & [0.0500,\,0.9997]
                           & 0.0105 & 0.0026       & [0.0003,\,0.1241] \\
    Miner Top 1\% Share    & 0.9311 & 0.9977       & [0.2665,\,0.99998]
                           & 0.1374 & 0.1160       & [0.0233,\,0.4838] \\[4pt]
    Validator Gini         & 0.9770 & 0.9813       & [0.8953,\,0.9926]
                           & 0.9789 & 0.9831       & [0.8999,\,0.9933] \\
    Validator HHI          & 0.0915 & 0.0600       & [0.0335,\,0.4708]
                           & 0.1141 & 0.0669       & [0.0185,\,0.6304] \\
    Validator Top 1\% Share& 0.6453 & 0.6648       & [0.3012,\,0.9816]
                           & 0.6802 & 0.6903       & [0.3253,\,0.9956] \\
    \bottomrule
  \end{tabular}
  \caption{Summary of stake and reward concentration metrics across subnets. Range is [min,\,max]. Top 1\% Share is the fraction of stake/ reward held by the richest percentile.}
  \label{tab:concentration}
\end{table}

\subsection{Stake and Reward Concentration}

Figure~\ref{fig:top1_parallel} (left) shows how the fraction of total stake held by the top 1~\% of wallets varies across subnets. Here, we can see that the richest percentile of wallets controls between roughly 38~\% and nearly 100~\% of total stake, with a median of 90~\%. In fact, there are only three of the subnets (greyed out in the figure) where the top 1~\% controls less than 50~\% of the total stake. Combined with Gini$\approx0.98$ and HHI$\approx0.14$ shown in Table~\ref{tab:concentration}, this demonstrates extreme stake skew: only a handful of wallets can collude to achieve a majority control in most subnets. 

Similarly, the richest percentile of wallets captures 12~\%–56~\% of all rewards (median $\approx24$~\%), as depicted in Figure~\ref{fig:top1_parallel} (right).  Although slightly less concentrated than stake, reward shares remain heavily skewed (Reward Gini$\approx0.76$, Reward HHI$\approx0.06$). This suggests that top performers still monopolize rewards—posing an additional collusion vector.

\subsection{Size \& Maturity Effects}

We also confirm two robust cross‑subnet trends: Figure~\ref{fig:size_age_hhi} (left) shows the relationship between subnet size (measured by the number of unique wallets) and average HHI, while the right panel shows subnet age versus average HHI. In both cases, we observe a pronounced downward trend: as participation broadens or as a subnet matures, concentration decreases:

\begin{figure}[h]
  \centering
  \includegraphics[width=0.48\textwidth,trim=0 100 0 100,clip]{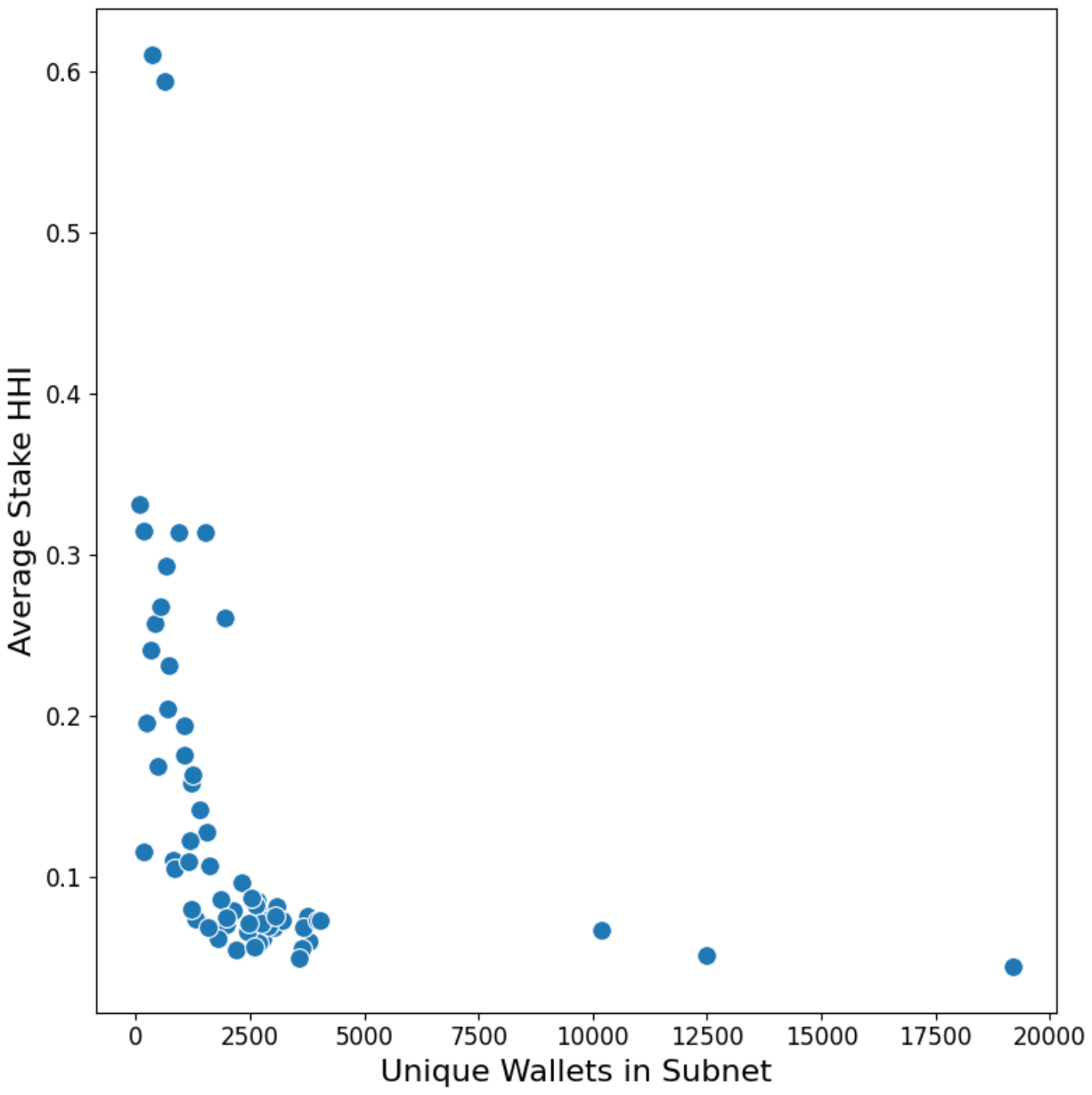}
  \hfill
  \includegraphics[width=0.48\textwidth,trim=0 100 0 100,clip]{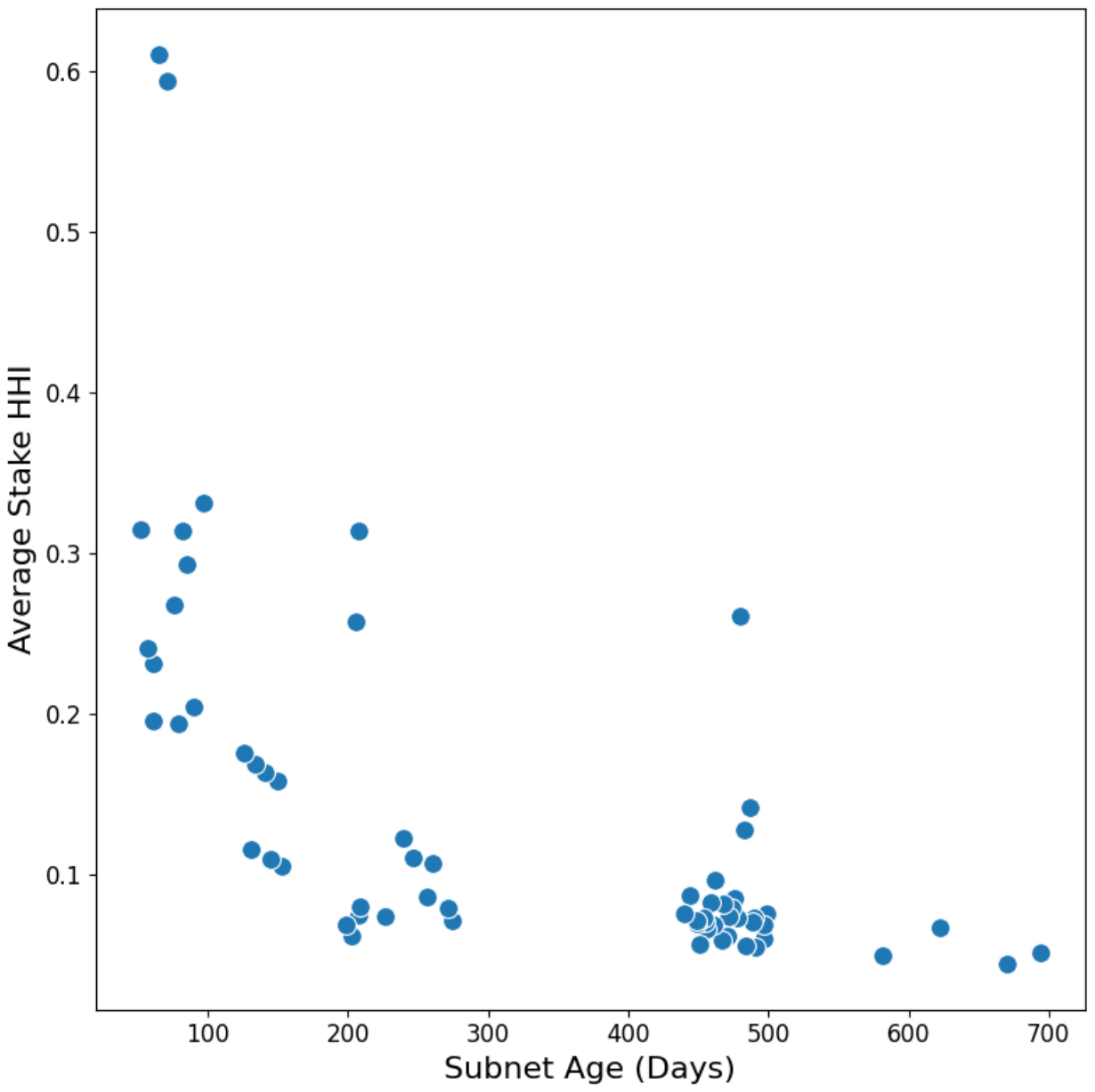}
  \caption{%
    \textbf{Left:} Average HHI vs.\ unique wallet count across subnets.  
    \textbf{Right:} Average HHI vs.\ subnet age.  
    Higher HHI signals a more concentrated market. Both panels exhibit a clear downward slope, indicating that larger or older subnets are less concentrated.%
  }
  \label{fig:size_age_hhi}
\end{figure}

\begin{itemize}
  \item \textbf{Unique Wallets vs.\ HHI.} Subnets with more participants have lower HHI (Pearson $r=-0.67$, $p<0.001$).  A diffuse stake base makes 51\% collusion more costly.
  \item \textbf{Subnet Age vs.\ HHI.} Older subnets show declining HHI (Pearson $r=-0.59$, $p<0.001$), plateauing near 0.08 after six months.  Maturation naturally diffuses early adopter dominance.
\end{itemize}

Together, these findings confirm that \emph{both scale and maturity drive decentralization}.  Large, well‑established subnets exhibit materially less concentration, while tiny or freshly launched subnets remain vulnerable to majority‐stake attacks.

\subsection{Stake/Reward/Performance Correlations}

\begin{figure}[h]
  \centering
  \includegraphics[width=0.48\textwidth]{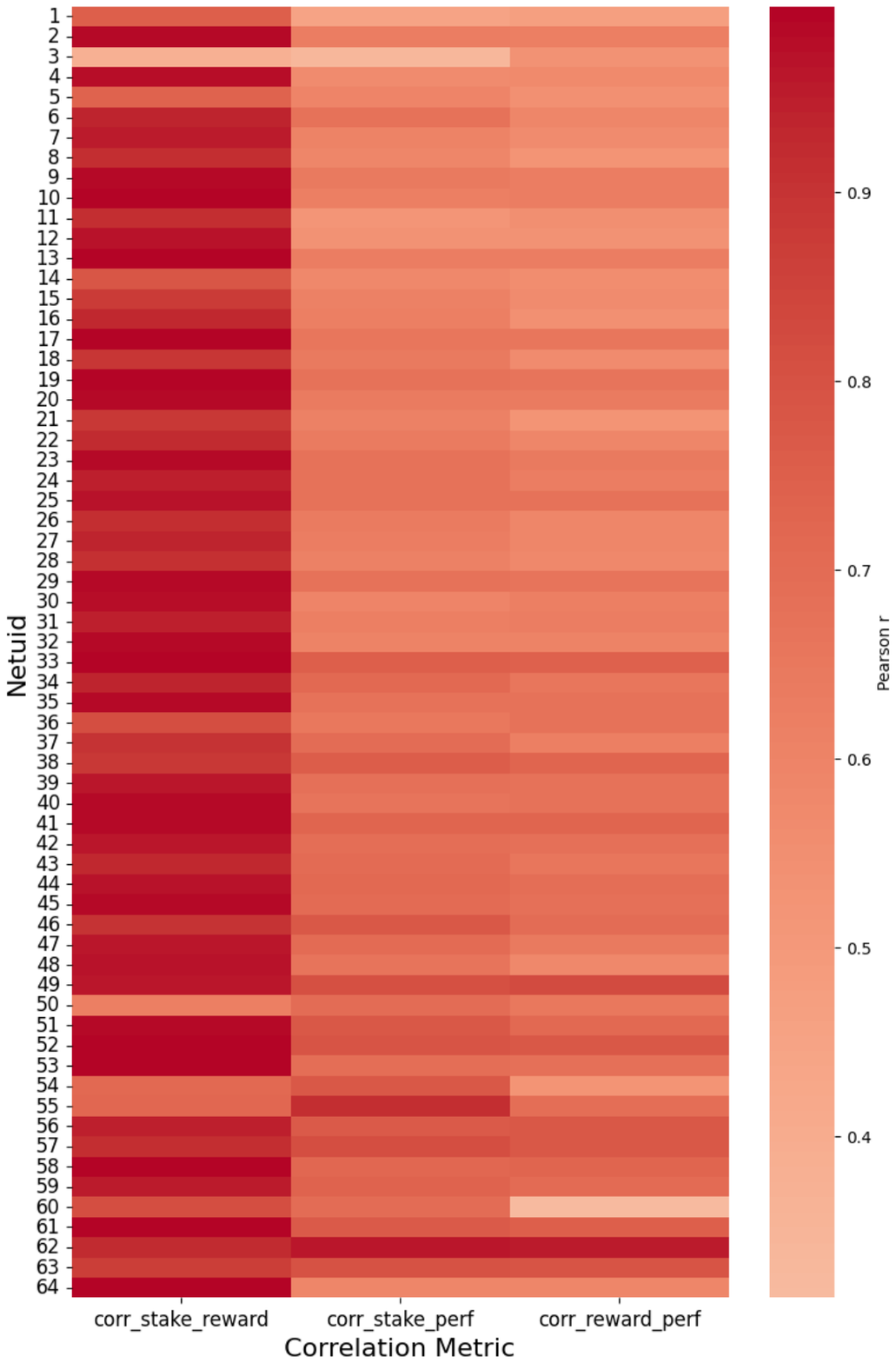}
  \hfill
  \includegraphics[width=0.48\textwidth]{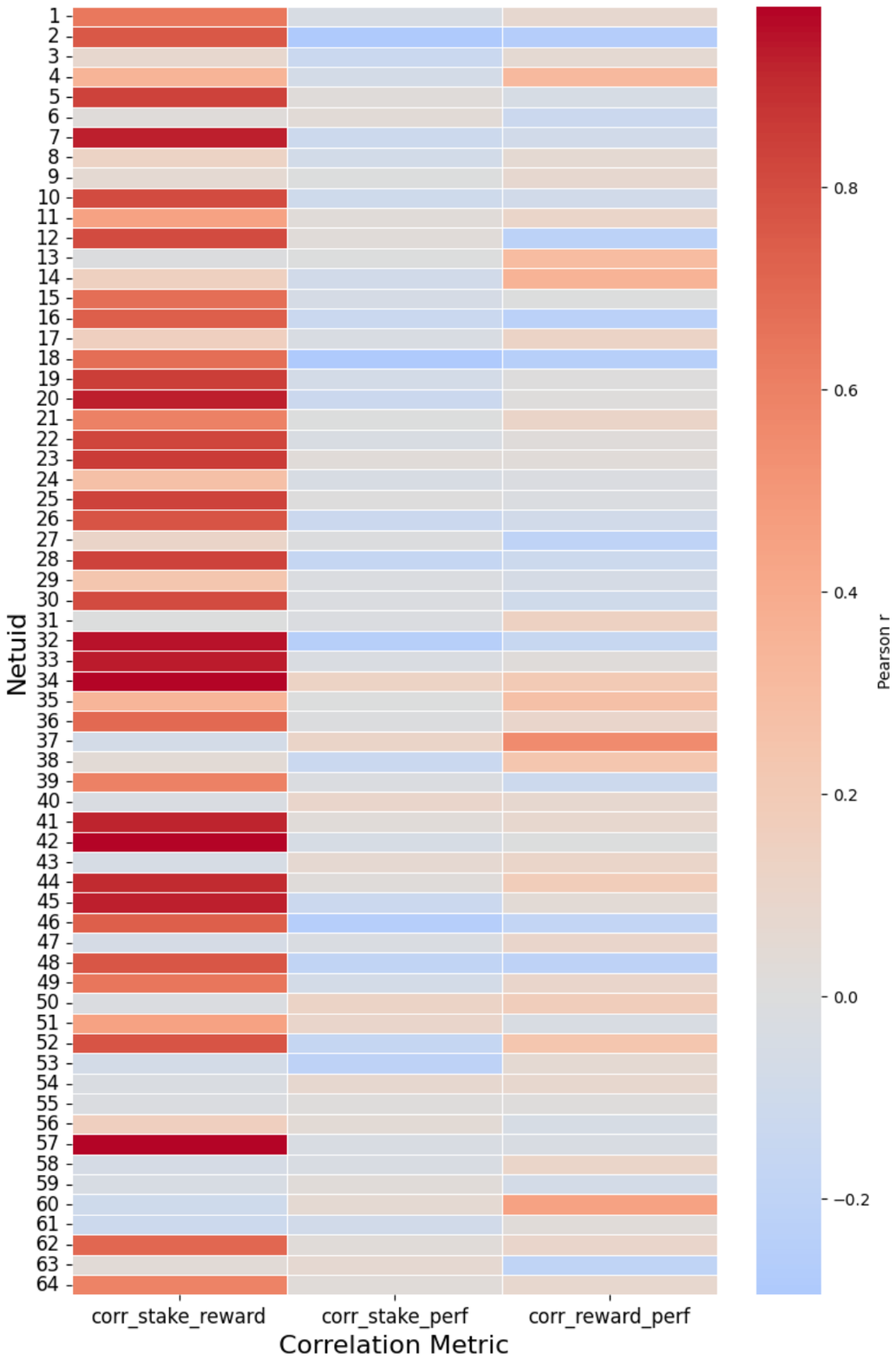}
  \caption{Per‐subnet Pearson correlations between stake, performance, and reward for validators (left) and miners (right).  Color intensity maps correlation strength (red= positive, blue=negative).}
  \label{fig:stake_perf_corr}
\end{figure}

 Note that the YC algorithm determines emission distributions within each subnet by combining a performance-based weight matrix with the amount of stake associated with each participant’s Unique Identifier (UID)\cite{bittensor_incentive}. In other words, rewards for miners and validators are, in principle, a function of both their staked TAO and their demonstrated performance on the network. To probe how economic power, earnings, and “quality” interact \emph{in practice}, we carried out the following analysis:

\begin{enumerate}
  \item \textbf{Aggregate performance metrics.} For each wallet we computed 
  \[
     \text{perf} = 
     \begin{cases}
       \text{trust}, & \text{if role = miner},\\
       \text{validator\_trust}, & \text{if role = validator}.
     \end{cases}
  \]
    Here, both fields are provided by the Bittensor API:
  \begin{itemize}
    \item \textbf{trust:} A score assigned by subnet validators reflecting how “trustworthy” a miner’s model outputs are, i.e.\ the degree to which other validators agree with its evaluations.
    \item \textbf{validator\_trust:} A consensus‐alignment score for subnet validators, indicating how closely each validator’s weight assignments align with the median of all validators in that subnet.
  \end{itemize}
  \item \textbf{Compute per‐subnet, per‐role correlations.} For each \((\text{netuid},\text{role})\) pair we measured three Pearson correlation coefficients:
  \[
    r_{s,r}=\mathrm{corr}(\text{stake},\,\text{reward}),\quad
    r_{s,p}=\mathrm{corr}(\text{stake},\,\text{perf}),\quad
    r_{p,r}=\mathrm{corr}(\text{perf},\,\text{reward}).
  \]
  \item \textbf{Visualize via heatmaps.} Figure~\ref{fig:stake_perf_corr} displays these three correlations for validators (left) and miners (right), with netuids on the vertical axis.
\end{enumerate}

We find that:
\begin{itemize}
  \item \textbf{Validators.}
    \begin{itemize}
      \item \(r_{s,r}\approx0.80\text{–}0.95\): High‐stake validators reliably secure larger reward shares.
      \item \(r_{s,p}\approx0.45\text{–}0.65\): Larger stake correlates moderately with consensus adherence.
      \item \(r_{p,r}\approx0.50\): Validators in stronger consensus earn somewhat higher rewards, though stake remains the dominant factor.
    \end{itemize}
  \item \textbf{Miners.}
    \begin{itemize}
      \item \(r_{s,r}\approx0.50\text{–}0.80\): Self‐bond magnitude drives server‐incentive earnings, but with broader dispersion than for validators.
      \item \(r_{s,p}\approx0\): Stake size is almost uncorrelated with “trust” scores, indicating that mere stake does not guarantee performance.
      \item \(r_{p,r}\approx0.10\text{–}0.30\): More trusted miners receive a slight rewards premium, but reward allocation remains heavily stake‐driven.
    \end{itemize}
\end{itemize}

These together imply that the very strong stake $\rightarrow$ reward link for both roles confirms that economic power translates directly into earnings.  However, the relatively weak performance $\rightarrow$ reward correlation—especially for miners—suggests an opportunity to rebalance incentives:  
\begin{itemize}
  \item \emph{Validators} could receive a larger share of rewards based on consensus quality (\(\text{validator\_trust}\)), further discouraging “weight copying.”  
  \item \emph{Miners} might be incentivized via an adjusted reward formula that more heavily weights \(\text{trust}\), thereby rewarding genuine utility over mere self‐bond.  
\end{itemize}

These insights provide a clear data‐driven roadmap for protocol refinements aimed at aligning economic incentives with honest, high‐quality participation.  

\subsection{The 51\% Attack \& Bittensor’s Current Vulnerability}

A “51\% attack” refers to any scenario in which a single actor or colluding coalition controls more than half of the resource that underpins consensus, enabling them to censor, reorder, or reverse transactions at will, or to stall finalization indefinitely. In Bitcoin this resource is hashing power: an attacker must amass over 50\% of the network’s total compute capacity—a capital‑ and energy‑intensive undertaking—to compromise the chain~\cite{gervais2016security,bonneau2015sok}. Network‑level manipulations such as eclipse attacks further illustrate how even temporary majority control can be leveraged for censorship or double‑spend vectors~\cite{heilman2015eclipse}. Bittensor’s consensus similarly rests on a scarce resource—staked TAO—but instead of hash‑rate, it is the fraction of total stake that grants voting power. When too few wallets hold a majority of stake, the barrier to a “51\% attack” collapses, since acquiring or coordinating these token‑holdings is far easier than mobilizing equivalent compute~\cite{gencer2018decentralization}.  

\begin{figure}[h]
  \centering
  \includegraphics[width=0.48\textwidth,trim=0 120 0 120,clip]{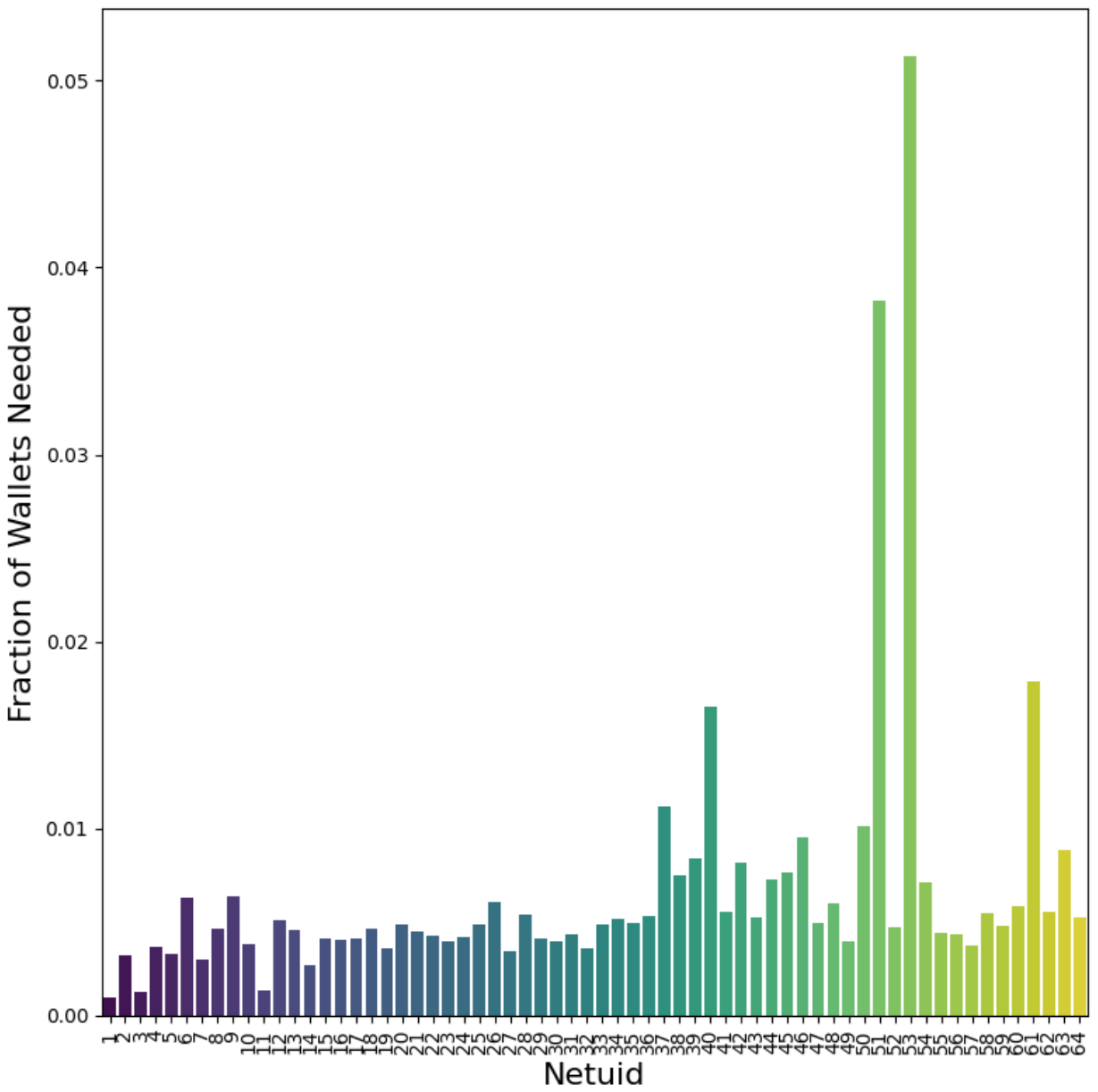}
  \hfill
  \includegraphics[width=0.48\textwidth,trim=0 120 0 120,clip]{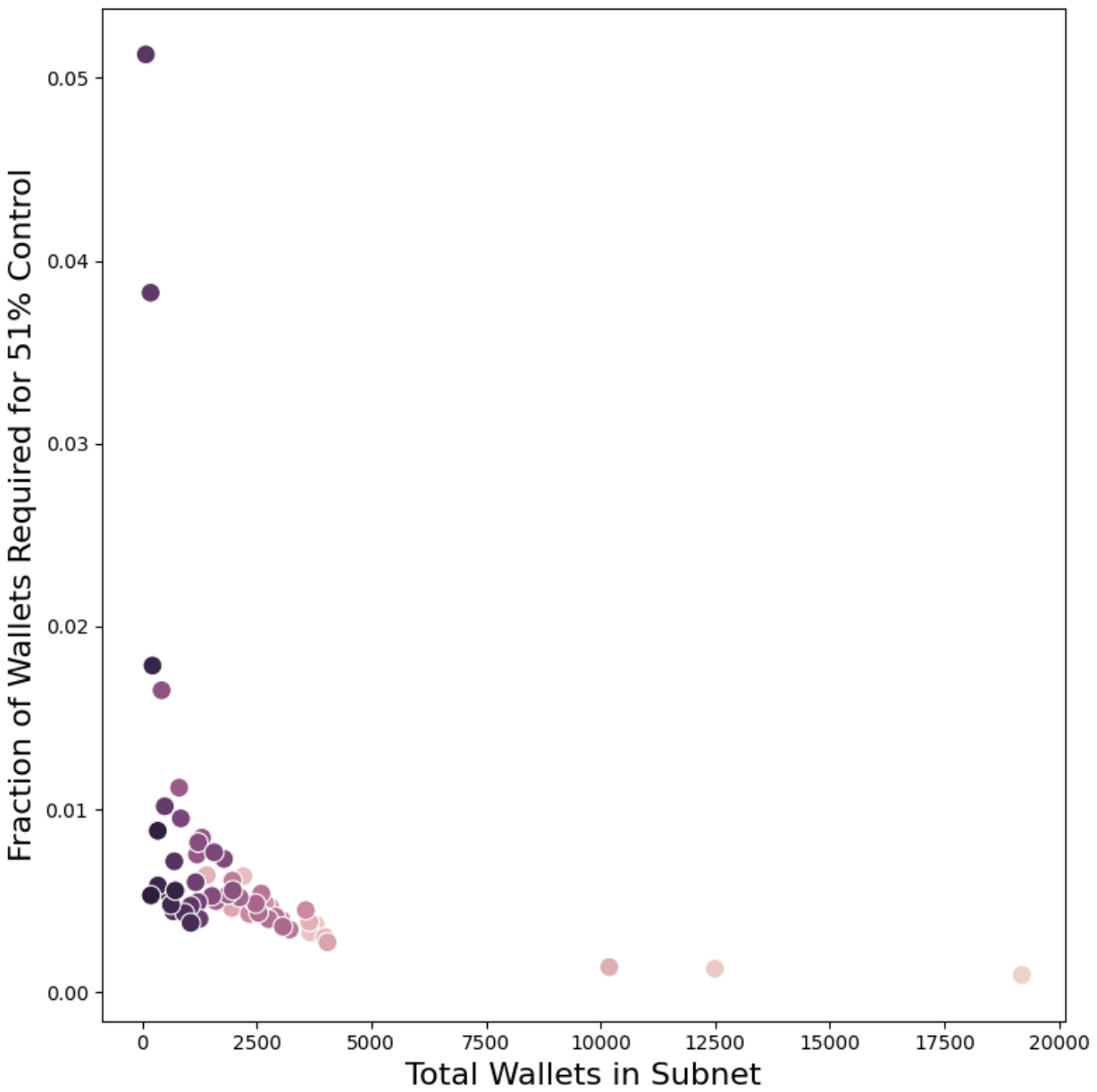}
  \caption{Left: Fraction of wallets required for majority (51\%) control by subnet, ordered by subnet ID. Right: Total unique wallets vs.\ fraction needed for 51\% control; marker color encodes the required fraction (darker=higher).}
  \label{fig:majority_control}
\end{figure}

Figure~\ref{fig:majority_control} quantifies how many wallets are needed to reach 51\% of total stake in each active subnet. We observe:

\begin{enumerate}
  \item \textbf{Extremely Low Bar in Many Subnets.}  
    Over half of subnets require \emph{under} 1\% of wallets to amass 51\% of stake (Figure~\ref{fig:majority_control}, left). In such cases, only a handful of addresses need to collude.

  \item \textbf{Size and Dispersion Drive Security.}  
    The right panel shows that mid‑sized subnets (a few hundred to a few thousand wallets) typically need around 1–2\,\% of wallets to collude. Very small subnets (a few hundred or fewer) can be compromised with even less—often <0.5\,\%.

  \item \textbf{Outliers Persist at All Scales.}  
    A few very tiny testnets spike toward 5\,\% because stake is fairly evenly split among very few participants. Conversely, some \emph{large} subnets (10K–20K wallets) still require <1\,\% when a handful of “whales” hold most of the stake.
\end{enumerate}

Taken together, under the current stake‑distribution regime:

\begin{itemize}
  \item \emph{Emerging or niche subnets} with low wallet counts are \emph{trivially} attackable by tiny coalitions.  
  \item \emph{Mid‑sized subnets} generally need on the order of 1–2\,\% of wallets to collude.  
  \item \emph{Largest subnets} only become harder to attack if stake is broadly dispersed; otherwise they too can be dominated by a few large holders (often <1\,\% of wallets).
\end{itemize}

\section{Proposed Solutions and Discussion}

\subsection{Realignment of Incentives}

\subsubsection{Performance‐Weighted Emission Split}

To more directly reward high‐quality contributions, we modify the validator-miner emission split parameter \(\xi\) in Bittensor's reward formula to depend linearly on each participant’s performance score.  Letting \(\xi_0\) be the base validator share (e.g.\ \(0.25\)) and \(\delta\) a sensitivity parameter, we set
\[
  \xi_i \;=\; \xi_0 \;+\; \delta\,\mathrm{perf}_i,
\]
and assign each wallet \(i\) an adjusted reward
\[
  r_i' =
  \begin{cases}
    r_i \times \bigl(\xi_0 + \delta\,\mathrm{perf}_i\bigr), 
      & i\text{ is a validator},\\[6pt]
    r_i \times \bigl((1-\xi_0) + \delta\,\mathrm{perf}_i\bigr), 
      & i\text{ is a miner},
  \end{cases}
\]
where \(r_i\) is the original stake‐based reward.  We then recompute, for each \((\mathrm{netuid},\mathrm{role})\), the Pearson correlations
\[
  r_{s,r} \;=\;\mathrm{corr}(\text{stake},\,\text{reward}), 
  \quad
  r_{p,r} \;=\;\mathrm{corr}(\text{perf},\,\text{reward})
\]
both at baseline (\(\delta=0\)) and under the adjusted rewards.

\paragraph{Implementation \& Results}  
Using our full dataset, we fix \(\xi_0=0.25\) and sweep \(\delta\) from 0 to 2.  At \(\delta=1\), miner correlations shift by
\[
  \Delta r_{s,r}\approx -0.018,\quad
  \Delta r_{p,r}\approx +0.032,
\]
while validator correlations change by only \(\sim-0.003\).  The complete sweep exhibits a steadily rising \(r_{p,r}\) for miners (up to \(\approx+0.04\) at \(\delta\approx1\)) and a modest erosion of \(r_{s,r}\) (down to \(-0.02\)), with negligible impact on validators.

\paragraph{Interpretation}  
This simple performance‐weighted split boosts rewards for more trusted miners—raising \(r_{p,r}\) by a few hundredths—at the cost of a slight weakening of the pure stake → reward link.  Because even \(\delta=1\) produces only a modest increase, more aggressive schemes (e.g.\ composite scoring or direct bonus multipliers) may be required to achieve stronger performance‐to‐reward alignment.

\subsubsection{Composite Scoring}

Instead of relying solely on the miner’s existing YC rank, we redefine each miner’s “effective rank” as a convex combination of that baseline rank \(R_j\) and their trust‐score:
\[
  R_j' \;=\; \lambda\,R_j \;+\; (1-\lambda)\,\mathrm{perf}_j,
  \quad \lambda\in[0,1].
\]
Miner emissions are then allocated in proportion to \(R_j'\), while validator rewards remain unchanged.  We sweep \(\lambda\) from 1 (pure stake) down to 0 (pure trust) and recompute, as before, the pairwise
\(\;r_{s,r}=\mathrm{corr}(\text{stake},\,\text{reward}),\;
  r_{p,r}=\mathrm{corr}(\text{perf},\,\text{reward})\)
for each \((\mathrm{netuid},\mathrm{role})\).

\paragraph{Implementation \& Results}  
In our full dataset we vary \(\lambda\in\{0,0.1,\dots,1.0\}\).  At \(\lambda=0.8\), for miners we observe
\[
  \Delta r_{s,r}\approx -0.10,\quad
  \Delta r_{p,r}\approx +0.035,
\]
and as \(\lambda\to0\) perf→reward climbs up to $\approx+0.36$ while stake→reward collapses toward $\approx0.05$.  Validators’ correlations shift by at most ±0.02 across the sweep.  

\paragraph{Interpretation}  
Composite scoring allows a strong rerouting of emissions toward high‑trust miners (up to a +0.36 boost), but at the expense of nearly destroying the stake incentive (stake→reward falls by $\approx0.90$).  It is thus a powerful but blunt instrument.

\subsubsection{Performance Bonus Multiplier}

Here we leave the 41$\colon$41$\colon$18 split intact and simply apply a small multiplicative “trust bonus” to each reward:
\[
  r_i^* = r_i \;\times\;\bigl(1 + \gamma\,\mathrm{perf}_i\bigr),
  \quad \gamma\ge0.
\]
We sweep \(\gamma\) from 0 to 0.2 and again recompute \(\{r_{s,r},r_{p,r}\}\).

\paragraph{Implementation \& Results}  
Over \(\gamma\in[0,0.20]\), miner perf→reward rises linearly up to $\approx+0.009$ at \(\gamma=0.2\), while stake→reward dips only modestly to $\approx–0.0035$. Validator correlations remain essentially flat (changes <0.001).

\paragraph{Interpretation}  
The bonus multiplier is the gentlest of the three tweaks: even a 20\% bonus yields only a +0.9pp lift in perf→reward while preserving >99\% of the stake→reward link.  It’s ideal for incremental, low‑risk governance experiments.

\subsubsection{Comparison of All Three Schemes}

\begin{figure}[h]
  \centering
  \includegraphics[width=\textwidth,trim=0 140 0 120,clip]{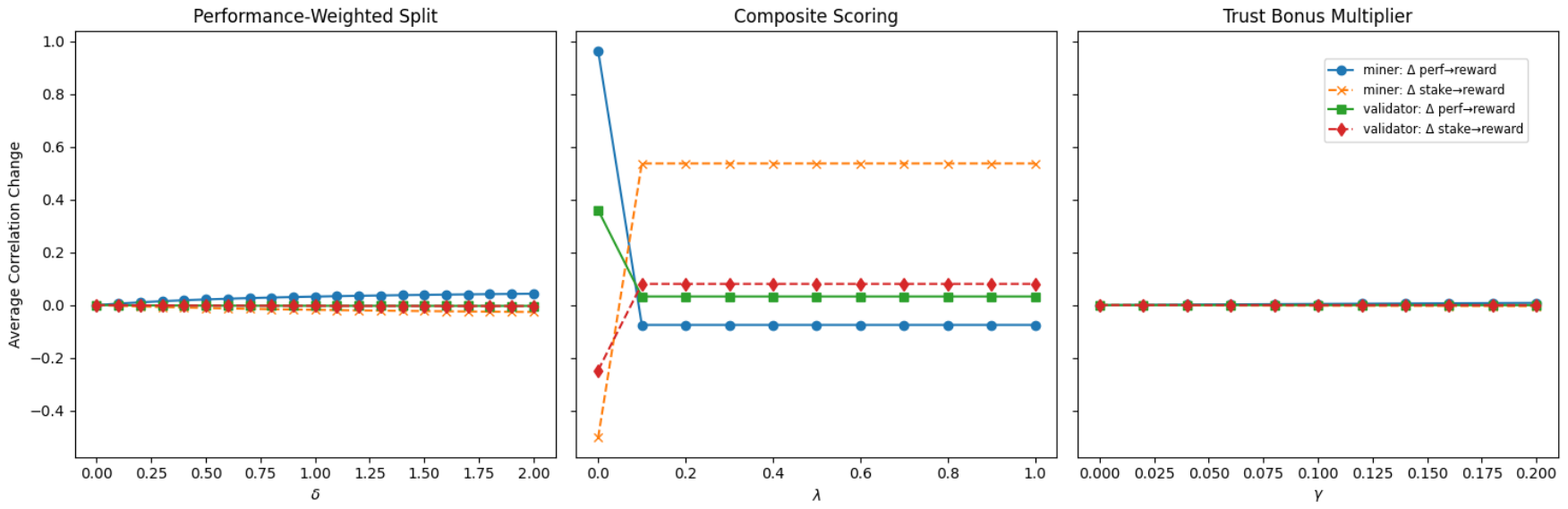}
  \caption{Correlation shifts under the three schemes: (a) performance‐weighted split, (b) composite scoring, (c) trust bonus. All panels share identical axes for direct visual comparison.}
  \label{fig:three_panels}
\end{figure}

Table~\ref{tab:scheme_summary} and Figure~\ref{fig:three_panels} summarize the trade‑offs:

\begin{table}[h]
  \centering
  \small
  \begin{tabular}{lccc}
    \toprule
     & \textbf{Perf Split} & \textbf{Composite} & \textbf{Bonus} \\
    \midrule
    Max \(\Delta r_{p,r}^{\text{miner}}\) 
      & +0.04          & +0.36        & +0.009        \\
    Min \(\Delta r_{s,r}^{\text{miner}}\) 
      & –0.02          & –0.91        & –0.0035       \\
    Validator impact 
      & $\leq0.003$        & $\leq0.02$       & $\leq0.001$       \\
    \bottomrule
  \end{tabular}
  \caption{Peak shifts in miner perf→reward and stake→reward correlations under each scheme.}
  \label{tab:scheme_summary}
\end{table}

\noindent
Figure~\ref{fig:three_panels} presents side–by–side comparisons of how each scheme shifts the key miner correlations $r_{s,r}$ (stake→reward) and $r_{p,r}$ (perf→reward).  In panel (a), the performance‑weighted split steadily increases $r_{p,r}$ by up to $\approx+0.04$ while only modestly reducing $r_{s,r}$ to $\approx–0.02$.  Panel(b) demonstrates the composite scoring’s extreme power: as $\lambda$ decreases, $r_{p,r}$ surges to $\approx+0.36$ but $r_{s,r}$ collapses by almost 0.90—effectively undermining any stake incentive.  Finally, panel(c) shows the trust‑bonus multiplier’s gentle touch: even at a 20~\% bonus, $r_{p,r}$ rises by under +0.01 and $r_{s,r}$ falls only $\approx–0.0035$.  Together these three panels visually confirm the trade‑off summarized in Table~\ref{tab:scheme_summary}: composite scoring yields the largest performance boost at the cost of virtually erasing the stake link, the performance‑weighted split strikes an intermediate balance, and the simple bonus multiplier gently nibbles at stake dominance while still favoring higher‐trust actors.

In short, all three approaches can raise the linkage between performance and payout. The composite method is the most powerful but almost eradicates the stake‐based incentive; the performance‐weighted split provides a middle ground (a few hundredths boost at $\approx–0.02$ stake penalty); and the bonus multiplier is the most conservative, nudging perf→reward by under a percentage point while barely denting the stake$\rightarrow$ reward relationship.

\subsection{Mitigating 51\% Attack Vulnerability}

To harden Bittensor against 51\% attacks, we compared three  protocol‑level ways to reshape stake (and thus reward) distributions:

\begin{itemize}
  \item \textbf{Stake Cap:} truncate any wallet’s effective stake above a chosen percentile (50~\%–99~\%).  
  \item \textbf{Nonlinear Weighting:} apply a concave transform \(s^\alpha\) (with \(\alpha=1.0, 0.9, \dots, 0.5\)) to flatten large stakes.  
  \item \textbf{Log‐Transform:} replace each stake \(s\) with \(\log(1+s)\), a natural diminishing‐returns curve.  
\end{itemize}

\subsubsection{1. Security vs.\ Whale‑Penalty Trade‑off.}  
\begin{figure}[h]
  \centering
  \includegraphics[width=0.48\textwidth,trim=0 100 0 100,clip]{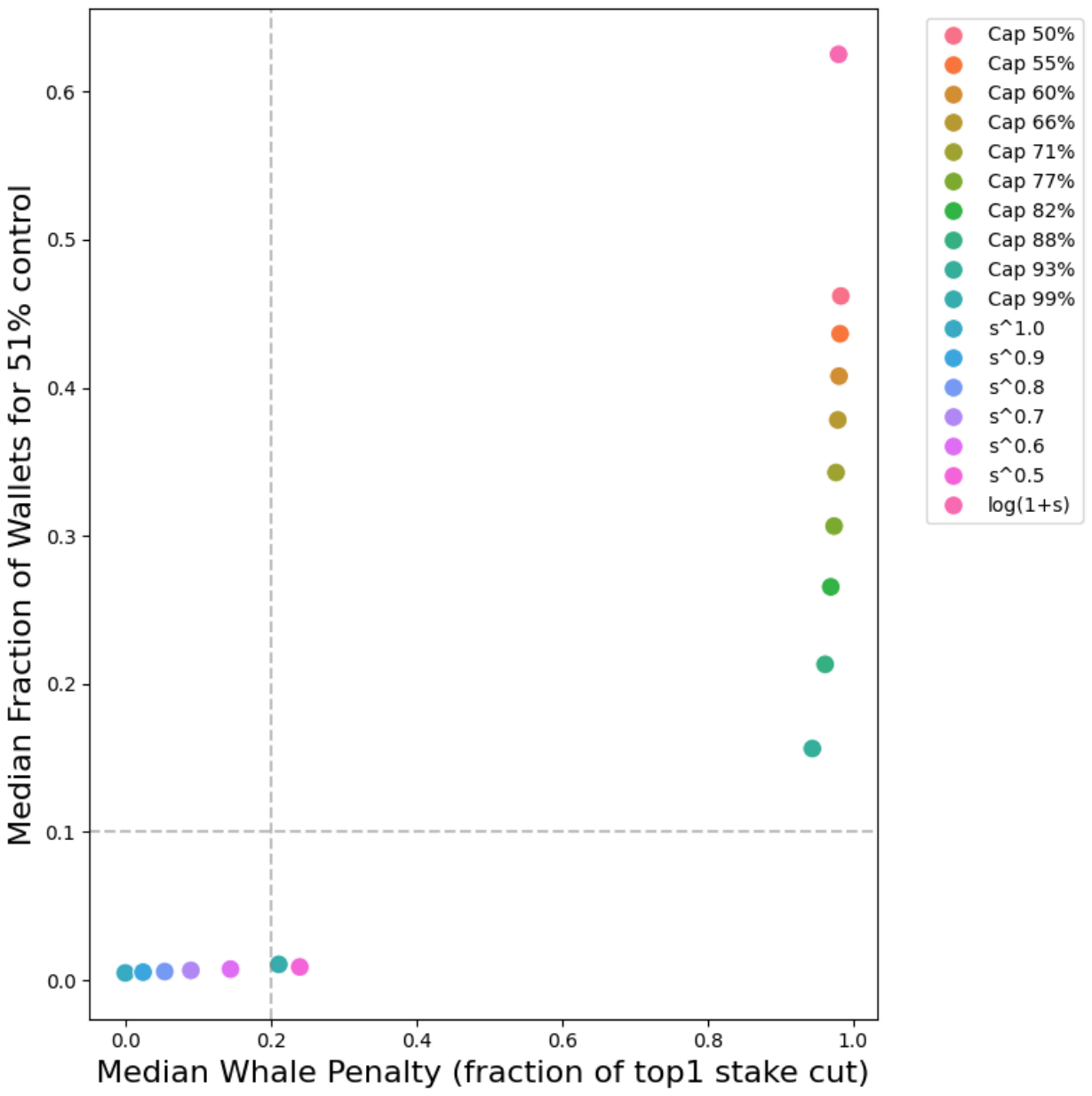}
  \hfill
  \includegraphics[width=0.48\textwidth,trim=0 100 0 100,clip]{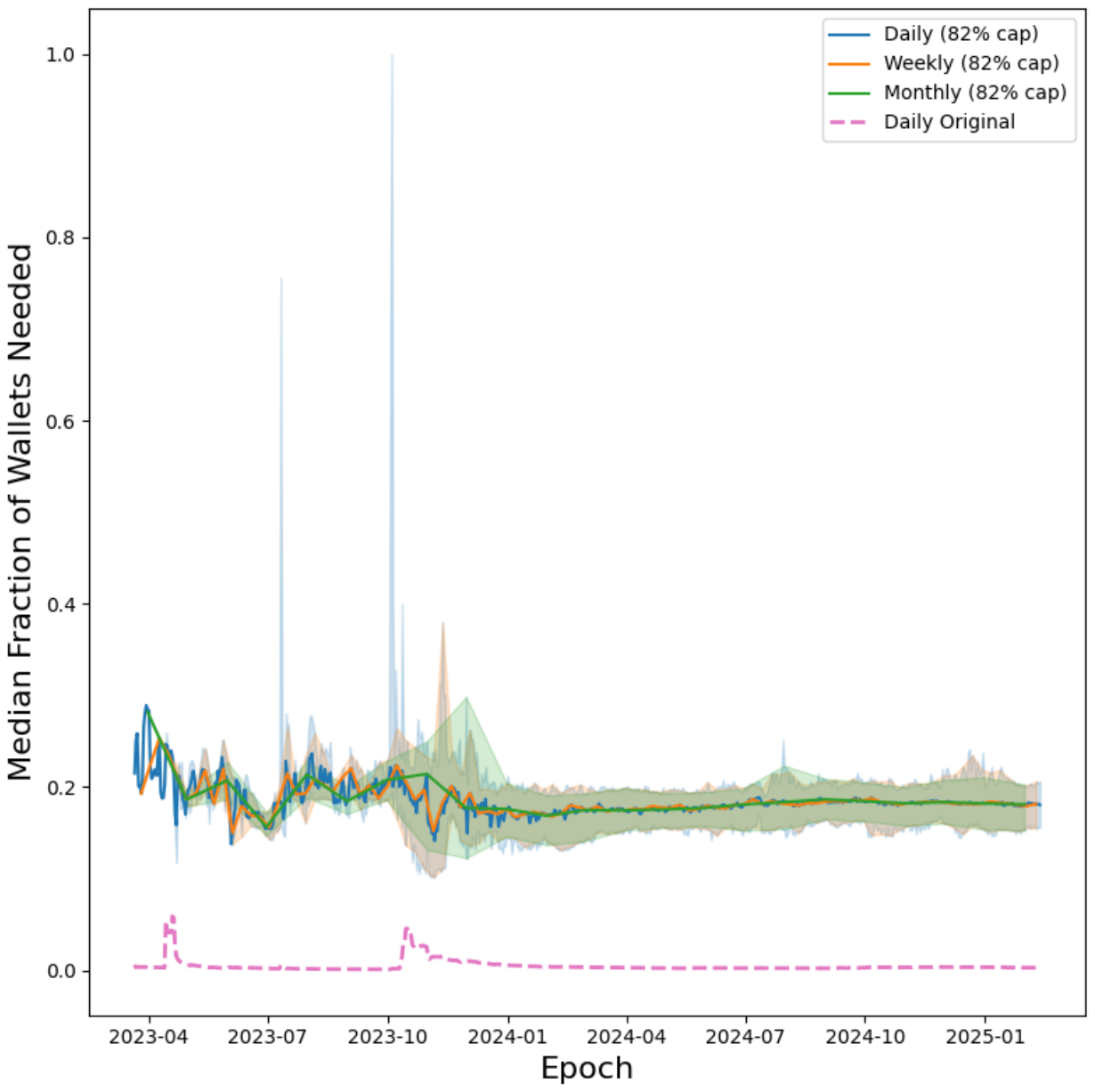}
  \caption{Left: Trade‑off between security (required coalition size) and whale‑penalty for various cap percentiles, highlighting the 88\% cap. Right: Median wallet‑fraction needed for 51\% control under an 88\% stake cap at daily (blue), weekly (orange), and monthly (green) snapshots, with shaded 10th–90th percentile bands and the uncapped baseline (dashed pink).}
  \label{fig:tradeoff}
\end{figure}

In Figure~\ref{fig:tradeoff} (left) we sweep each intervention over a range of parameters and plot:
\[
  \underbrace{\tilde f}_{\substack{\text{median wallet}\\\text{fraction for 51\%}}}
  \quad\text{vs.}\quad
  \underbrace{\Delta w}_{\substack{\text{median “whale‐penalty”:}\\\text{fraction of top‑1\% stake trimmed}}}.
\]
The “Cap” points (pink‐to‐teal) show percentiles from 50 \% up to 99 \%, the solid circles (blue‐to‐purple) show \(s^\alpha\) with \(\alpha=1.0\) down to 0.5, and the magenta circle shows \(\log(1+s)\).  A good intervention lives up and to the left: high security with low penalty.  Notice that the 88~\% cap (dark green) sits near the Pareto‐frontier.

\subsubsection{2. Temporal Robustness.}  
We then applied the winning 88~\% cap across daily, weekly and monthly snapshots and computed, at each timestep, the median wallet‐fraction for 51~\% control (with 10–90~\% bands).  Figure~\ref{fig:tradeoff} (right) shows that after a brief settling period, all three capped series converge around ~20~\%, with very tight dispersion—whereas the uncapped baseline remains stuck near 1–2~\%.  This confirms that the 88~\% cap is not only powerful, but also stable under arbitrary sampling frequencies.

\noindent\textbf{Summary.}  
All three distribution interventions can raise the 51~\% threshold dramatically compared to the uncapped baseline.  The concave \(s^\alpha\) and \(\log(1+s)\) transforms offer very low whale‐penalties but plateau near only a 2–3× security gain, whereas the 88~\% cap achieves a 20× gain with an acceptable ~22~\% whale‐penalty and exhibits strong temporal stability.  This makes the 88~\% cap our recommended first‐cut defense, with nonlinear weighting or log‐demurrage as lighter‐touch alternatives for incremental governance trials.

\section{Conclusion and Future Work}

We presented the first large‐scale empirical study of Bittensor’s on‐chain data, benchmarking its tokenomics and decentralization against Bitcoin.  Our analysis uncovered severe stake and reward concentration—fewer than 2~\% of wallets can command a 51~\% majority in most subnets. Further, we find that across all subnets, economic stake remains the dominant predictor of rewards, whereas performance contributes only modestly for validators and very weakly for miners, revealing a misalignment between “quality” and payout. To address that misalignment, we introduced three incentive‐realignment schemes—performance‐weighted emission split, composite scoring, and trust‐bonus multipliers—to more directly tie rewards to quality.  On the other hand, to remedy the broader security vulnerability, we not only evaluated three protocol‐level defenses against 51~\% attacks (stake caps, concave weighting, reward demurrage) but also proposed and tested incentive realignment schemes—performance‐weighted split, composite scoring, and trust‐bonus multipliers—that meaningfully boost the perf→reward linkage with controllable impacts on the stake→reward relationship.  

Looking ahead, assessing the impact of the recent dTAO upgrade on subnet valuation and security is a pressing empirical task.  Incorporating delegation dynamics—how stakers shift TAO among validators—will further illuminate power flows and potential centralization vectors.  Finally, developing a unified game‐theoretic model of miners, validators, and delegators under these enhanced incentives could sharpen our understanding of equilibrium behaviors and guide the design of truly Bitcoin‐grade decentralized AI markets.

\section*{Acknowledgements}
We thank Zhipeng Wang for his valuable guidance and insightful discussions. We also thank all anonymous reviewers for their helpful feedback. This work was partially supported by a research grant from the Ethereum Foundation.

\bibliographystyle{plain}

\clearpage
\appendix
\renewcommand{\thesection}{Appendix~\Alph{section}}

\section{Detailed Reward Mechanism (Pre-dTAO)}
\label{appendix:reward}

\subsection*{1. Emission Extraction Per Tempo}

At the end of each tempo (360 blocks) the emitted TAO is split:
\[
\text{18\% Subnet Owner} \quad\mid\quad
\text{41\% Miners} \quad\mid\quad
\text{41\% Validators \& Stakers}
\]
\cite{emissions_overview}

\subsection*{2. Validators}

Validators earn from the 41 \% validator bucket by bonding their stake to miners whose work they judge valuable. Each epoch they publish weights for miners; only the portions of those weights that exceed the stake-weighted consensus ceiling are clipped and penalised before being turned into time-smoothed bonds. The more stake a validator controls—and the more its long-term weights align with consensus—the larger its share of emissions.

\paragraph*{(i) Penalty-adjusted bond weight}
\[
\tilde{W}_{ij}= (1-\beta)\,W_{ij} + \beta\,\bar{W}_{ij}
\]\cite{yuma_consensus}
in which \(W_{ij}\) is the raw weight that validator \(i\) assigns to miner \(j\), \(\bar{W}_{ij}\) is the subnet-wide clipped weight (see Miner section), \(\beta\!\in[0,1]\) is the penalty factor, whereas \(\tilde{W}_{ij}\) is the resulting “bond weight’’ after out-of-consensus punishment. Out-of-consensus evaluations are diluted toward the consensus ceiling; accurate or conservative weights pass unchanged.

\smallskip

\paragraph*{(ii) Instant bond}
\[
\Delta B_{ij}= 
\frac{S_i\,\tilde{W}_{ij}}
     {\displaystyle\sum_{k\in V} S_k\,\tilde{W}_{kj}}
\]\cite{yuma_consensus}
in which \(S_i\) denotes the total stake (self + delegations) behind validator \(i\), \(V\) is the set of all validators, and \(\Delta B_{ij}\) is the fraction of \(S_i\) instantly “bonded’’ to miner \(j\).  Each validator splits its stake across miners in proportion to its penalty-adjusted weights; all fractions sum to 1.

\smallskip

\paragraph*{(iii) EMA bond update}
\[
B_{ij}(t)=\alpha\,\Delta B_{ij} + (1-\alpha)\,B_{ij}(t-1)
\]\cite{yuma_consensus}
in which \(\alpha\) is an EMA smoothing parameter and \(B_{ij}(t)\) the running bond at tempo \(t\).  Bonds build up (or decay) gradually, rewarding validators who stay aligned with consensus over time.

\smallskip

\paragraph*{(iv) Validator emission share}
\[
V_i=\sum_{j\in M} B_{ij}\,M_j
\]\cite{yuma_consensus}
in which \(M\) is the set of miners, \(M_j\) is miner \(j\)’s emission share (next subsection), and \(V_i\) is validator \(i\)’s slice of the 41 \% validator pool. Validators earn most when their long-term bonds point to miners the subnet collectively rewards.

\paragraph*{Reward Drivers}\vspace{-0.4em}
\begin{itemize}
  \item Stake size \(S_i\)
  \item Alignment with consensus (low \(\beta\) penalty)
  \item Smoothing factor \(\alpha\)
  \item Quality of bonded miners (\(M_j\))
\end{itemize}

\subsection*{3. Miners}

Miners compete for the 41 \% miner bucket. Every validator scores each miner, but any score above the stake-weighted median ceiling is clipped away; the remaining stake-weighted scores are summed into an aggregate ranking. A miner’s reward is simply its share of that ranking relative to all miners in the subnet, so broad support from high-stake validators — not a few inflated ratings — is what maximises payout.

\paragraph*{(i) Stake-weighted clipping benchmark}
\[
\bar{W}_{j}= \arg\max_{w}
          \Bigl(\!
          \sum_{i\in V} S_i\,\mathbf{1}[W_{ij}\ge w]\;\ge\;\kappa
          \Bigr)
\]\cite{yuma_consensus}
in which \(W_{ij}\) is the raw weight from validator \(i\) to miner \(j\), \(S_i\) the validator’s stake, \(\kappa\!(=0.5\text{ by default})\) the clipping threshold, whereas \(\bar{W}_{j}\) is the highest score backed by at least \(\kappa\) fraction of stake. The median-stake validator sets an upper limit on how much praise a miner can receive.

\smallskip

\paragraph*{(ii) Clipped weight}
\[
\bar{W}_{ij}= \min\!\bigl(W_{ij},\,\bar{W}_{j}\bigr)
\]\cite{yuma_consensus}
in which \(\bar{W}_{ij}\) is the post-clip weight actually recognised by the protocol. Any excess above the benchmark is discarded; neither miner nor validator gains from inflated scores.

\smallskip

\paragraph*{(iii) Aggregate ranking}
\[
R_j= \sum_{i\in V} S_i\,\bar{W}_{ij}
\]\cite{yuma_consensus}
in which \(R_j\) is the stake-weighted sum of post-clip weights for miner \(j\). Broad, stake-backed support drives the miner’s raw ranking.

\smallskip

\paragraph*{(iv) Miner emission share}
\[
M_j= \frac{R_j}{\displaystyle\sum_{k\in M} R_k}
\]\cite{yuma_consensus}
in which the denominator is the subnet’s total ranking mass, and \(M_j\) is miner \(j\)’s fraction of the fixed 41 \% miner bucket. A miner’s reward grows with its own \(R_j\) but shrinks if many other miners also achieve high rankings.

\paragraph*{Reward Drivers}\vspace{-0.4em}
\begin{itemize}
  \item Post-clip weights \(\bar{W}_{ij}\) from validators
  \item Stake of those validators \(S_i\)
  \item Clipping parameter \(\kappa\)
  \item Competition from other miners in the same subnet
\end{itemize}

\subsection*{4. Delegators}

Delegators (otherwise known as "nominators") lock their \$TAO behind a validator and receive a share of that validator’s emissions after the validator takes its commission. Thus a delegator’s payout depends on both their own stake and the performance/commission of the validator they choose. Commission is set by each validator. 

\paragraph*{(i) Delegator emission share}
\[
R_d = (1 - T_i)\,
      \frac{S_d}{S_i}\,
      D_i
\]\cite{staking_delegation}
in which \(T_i\) is the validator \(i\)’s commission (take) rate,
\(S_d\) is the amount of TAO delegated by delegator \(d\),
\(S_i\) is the total stake backing validator \(i\),
and \(D_i\) is the validator’s own reward before commission. The validator first removes its commission; the remaining emissions are split proportionally among delegators according to their share of the stake pool.

\paragraph*{Reward Drivers}\vspace{-0.4em}
\begin{itemize}
  \item Size of personal delegation \(S_d\) relative to the pool \(S_i\)
  \item Validator commission \(T_i\) (lower commission \(\rightarrow\) higher \(R_d\))
  \item Validator performance \(D_i\) (itself driven by stake, consensus
        alignment, and quality of bonded miners)
\end{itemize}

\clearpage
\section{Definitions of Gini Coefficient and Herfindahl–Hirschman Index}
\label{appendix:gini}

\subsection*{1. Gini Coefficient}

The Gini coefficient \(G\) measures \emph{inequality} in a distribution
(e.g.\ wealth, stake, emissions).  It ranges from \(0\) (perfect equality) to
\(1\) (perfect concentration).

\paragraph*{Lorenz‐curve definition}
\[
G = 1 - 2 \int_{0}^{1} L(p)\,dp
\]
in which \(L(p)\) is the Lorenz curve giving the cumulative share of the
resource held by the bottom \(p\) fraction of agents.  
\textit{Logic.}  The integral captures the area between perfect equality
(\(L(p)=p\)) and the actual distribution.

\smallskip

\paragraph*{Discrete form (sample of \(n\))}
\[
G = \frac{1}{2n^2\bar{x}}
    \sum_{i=1}^{n}\sum_{j=1}^{n}
    \lvert x_i - x_j \rvert
\]
in which \(x_i\) is the resource of agent \(i\) and
\(\bar{x}\) the sample mean.  
\textit{Logic.}  The double sum averages all absolute pairwise differences,
then normalises by population size and mean to lie in \([0,1]\).

\paragraph*{Interpretation}\vspace{-0.4em}
\begin{itemize}
  \item \(G\approx 0\): resource evenly spread.
  \item \(G\approx 1\): most of the resource in very few hands.
\end{itemize}

\subsection*{2. Herfindahl–Hirschman Index (HHI)}

The HHI gauges \emph{market concentration}.  For shares
\(s_i\) (expressed as fractions summing to \(1\)), it is the sum of squared
shares:

\[
\text{HHI} = \sum_{i=1}^{n} s_i^{\,2}
\]

in which \(s_i = x_i / \sum_k x_k\) is the share of agent \(i\).  
\textit{Logic.}  Squaring magnifies large shares, so dominance by a few agents
pushes the index upward.

\paragraph*{Scale and thresholds}\vspace{-0.4em}
\begin{itemize}
  \item \(\text{HHI}_{\min}=1/n\): perfectly equal shares.
  \item \(\text{HHI}_{\max}=1\): a single agent holds 100 
  \item Common antitrust cut-offs (U.S.\ DOJ):  
        \(\text{HHI}<0.15\) unconcentrated,  
        \(0.15\le\text{HHI}<0.25\) moderately concentrated,  
        \(\text{HHI}\ge0.25\) highly concentrated.
\end{itemize}

\end{document}